\documentclass[preprint,tightenlines,superscriptaddress,showpacs]{revtex4}
\usepackage{times}
\usepackage{amsmath}
\usepackage{graphicx}
\usepackage{epsfig}
\usepackage{dcolumn}
\usepackage{bm}
\newcommand{\BR}{{\cal B}}
\newcommand{\pp}{\pi^+\pi^-}

\newcommand{\kk}{K^+K^-}
\newcommand{\LL}{\ell^+\ell^-}
\newcommand{\EE}{e^+e^-}
\newcommand{\MM}{\mu^+\mu^-}

\newcommand{\psip}{\psi(2S)}

\newcommand{\jpsi}{J/\psi}
\newcommand{\ppjpsi}{\pi^+\pi^-J/\psi}

\newcommand{\zc}{Z(3900)^\pm}

\newcommand{\ksksjpsi}{K_S^0K_S^0J/\psi}
\newcommand{\kkjpsi}{K^+K^- J/\psi}

\begin{document}

\preprint{} \preprint{\vbox{ \hbox{   }
                        \hbox{Belle Preprint 2014-2}
                        \hbox{KEK   Preprint 2013-61}
                       }}

\title{\quad\\[1.cm] \boldmath
Updated Cross Section Measurement of $e^+ e^- \to K^+ K^- J/\psi$
and $K_S^0K_S^0J/\psi$ via Initial State Radiation at Belle}


\noaffiliation
\affiliation{University of the Basque Country UPV/EHU, 48080 Bilbao}
\affiliation{Beihang University, Beijing 100191}
\affiliation{University of Bonn, 53115 Bonn}
\affiliation{Budker Institute of Nuclear Physics SB RAS and Novosibirsk State University, Novosibirsk 630090}
\affiliation{Faculty of Mathematics and Physics, Charles University, 121 16 Prague}
\affiliation{University of Cincinnati, Cincinnati, Ohio 45221}
\affiliation{Deutsches Elektronen--Synchrotron, 22607 Hamburg}
\affiliation{Justus-Liebig-Universit\"at Gie\ss{}en, 35392 Gie\ss{}en}
\affiliation{II. Physikalisches Institut, Georg-August-Universit\"at G\"ottingen, 37073 G\"ottingen}
\affiliation{Gyeongsang National University, Chinju 660-701}
\affiliation{Hanyang University, Seoul 133-791}
\affiliation{University of Hawaii, Honolulu, Hawaii 96822}
\affiliation{High Energy Accelerator Research Organization (KEK), Tsukuba 305-0801}
\affiliation{Hiroshima Institute of Technology, Hiroshima 731-5193}
\affiliation{IKERBASQUE, Basque Foundation for Science, 48011 Bilbao}
\affiliation{Indian Institute of Technology Guwahati, Assam 781039}
\affiliation{Indian Institute of Technology Madras, Chennai 600036}
\affiliation{Institute of High Energy Physics, Chinese Academy of Sciences, Beijing 100049}
\affiliation{Institute of High Energy Physics, Vienna 1050}
\affiliation{Institute for High Energy Physics, Protvino 142281}
\affiliation{INFN - Sezione di Torino, 10125 Torino}
\affiliation{Institute for Theoretical and Experimental Physics, Moscow 117218}
\affiliation{J. Stefan Institute, 1000 Ljubljana}
\affiliation{Kanagawa University, Yokohama 221-8686}
\affiliation{Kennesaw State University, Kennesaw GA 30144}
\affiliation{Institut f\"ur Experimentelle Kernphysik, Karlsruher Institut f\"ur Technologie, 76131 Karlsruhe}
\affiliation{Department of Physics, Faculty of Science, King Abdulaziz University, Jeddah 21589}
\affiliation{Korea Institute of Science and Technology Information, Daejeon 305-806}
\affiliation{Korea University, Seoul 136-713}
\affiliation{Kyungpook National University, Daegu 702-701}
\affiliation{\'Ecole Polytechnique F\'ed\'erale de Lausanne (EPFL), Lausanne 1015}
\affiliation{Faculty of Mathematics and Physics, University of Ljubljana, 1000 Ljubljana}
\affiliation{Luther College, Decorah, Iowa 52101}
\affiliation{University of Maribor, 2000 Maribor}
\affiliation{Max-Planck-Institut f\"ur Physik, 80805 M\"unchen}
\affiliation{School of Physics, University of Melbourne, Victoria 3010}
\affiliation{Moscow Physical Engineering Institute, Moscow 115409}
\affiliation{Graduate School of Science, Nagoya University, Nagoya 464-8602}
\affiliation{Kobayashi-Maskawa Institute, Nagoya University, Nagoya 464-8602}
\affiliation{Nara Women's University, Nara 630-8506}
\affiliation{National Central University, Chung-li 32054}
\affiliation{National United University, Miao Li 36003}
\affiliation{Department of Physics, National Taiwan University, Taipei 10617}
\affiliation{H. Niewodniczanski Institute of Nuclear Physics, Krakow 31-342}
\affiliation{Nippon Dental University, Niigata 951-8580}
\affiliation{Niigata University, Niigata 950-2181}
\affiliation{University of Nova Gorica, 5000 Nova Gorica}
\affiliation{Osaka City University, Osaka 558-8585}
\affiliation{Pacific Northwest National Laboratory, Richland, Washington 99352}
\affiliation{Panjab University, Chandigarh 160014}
\affiliation{Peking University, Beijing 100871}
\affiliation{University of Science and Technology of China, Hefei 230026}
\affiliation{Seoul National University, Seoul 151-742}
\affiliation{Soongsil University, Seoul 156-743}
\affiliation{Sungkyunkwan University, Suwon 440-746}
\affiliation{School of Physics, University of Sydney, NSW 2006}
\affiliation{Department of Physics, Faculty of Science, University of Tabuk, Tabuk 71451}
\affiliation{Tata Institute of Fundamental Research, Mumbai 400005}
\affiliation{Excellence Cluster Universe, Technische Universit\"at M\"unchen, 85748 Garching}
\affiliation{Tohoku Gakuin University, Tagajo 985-8537}
\affiliation{Tohoku University, Sendai 980-8578}
\affiliation{Department of Physics, University of Tokyo, Tokyo 113-0033}
\affiliation{Tokyo Institute of Technology, Tokyo 152-8550}
\affiliation{Tokyo Metropolitan University, Tokyo 192-0397}
\affiliation{Tokyo University of Agriculture and Technology, Tokyo 184-8588}
\affiliation{University of Torino, 10124 Torino}
\affiliation{CNP, Virginia Polytechnic Institute and State University, Blacksburg, Virginia 24061}
\affiliation{Wayne State University, Detroit, Michigan 48202}
\affiliation{Yamagata University, Yamagata 990-8560}
\affiliation{Yonsei University, Seoul 120-749}
  \author{C.~P.~Shen}\affiliation{Beihang University, Beijing 100191} 
  \author{C.~Z.~Yuan}\affiliation{Institute of High Energy Physics, Chinese Academy of Sciences, Beijing 100049} 
  \author{P.~Wang}\affiliation{Institute of High Energy Physics, Chinese Academy of Sciences, Beijing 100049} 
  \author{A.~Abdesselam}\affiliation{Department of Physics, Faculty of Science, University of Tabuk, Tabuk 71451} 
  \author{I.~Adachi}\affiliation{High Energy Accelerator Research Organization (KEK), Tsukuba 305-0801} 
  \author{H.~Aihara}\affiliation{Department of Physics, University of Tokyo, Tokyo 113-0033} 
  \author{S.~Al~Said}\affiliation{Department of Physics, Faculty of Science, University of Tabuk, Tabuk 71451}\affiliation{Department of Physics, Faculty of Science, King Abdulaziz University, Jeddah 21589}
  \author{D.~M.~Asner}\affiliation{Pacific Northwest National Laboratory, Richland, Washington 99352} 
  \author{V.~Aulchenko}\affiliation{Budker Institute of Nuclear Physics SB RAS and Novosibirsk State University, Novosibirsk 630090} 
  \author{T.~Aushev}\affiliation{Institute for Theoretical and Experimental Physics, Moscow 117218} 
% ...
\collaboration{The Belle Collaboration}

\begin{abstract}

The cross sections of the processes $\EE\to \kkjpsi$ and
$\ksksjpsi$ are measured via initial state radiation at
center-of-mass energies between the threshold and 6.0~GeV using a
data sample of 980~fb$^{-1}$ collected with the Belle detector on
or near the $\Upsilon(nS)$ resonances, where $n=$1, 2, ..., 5. The
cross sections for $\EE\to \kkjpsi$ are at a few pb level and the
average cross section for $\EE \to \ksksjpsi$ is $1.8\pm 0.6 (\rm
stat.)\pm 0.3 (\rm syst.)$~pb between 4.4 and 5.2~GeV. All of them
are consistent with previously published results with improved
precision. A search for resonant structures and associated
intermediate states in the cross section of the process $\EE\to
\kkjpsi$ is performed.

\end{abstract}

\pacs{14.40.Gx, 13.25.Gv, 13.66.Bc}

\maketitle


\section{Introduction}

Recently a new charged charmoniumlike state, the $\zc$, was
observed by the Belle~\cite{Belle-z} and  BESIII~\cite{bes3-zc}
experiments in a study of $\EE \to \ppjpsi$ at center-of-mass (CM)
energies around 4.26~GeV. It was soon confirmed  with the CLEO
data at a CM energy of 4.17 GeV~\cite{seth}. As the $\zc$ state
has a strong coupling to charmonium and is charged, it can not be
a conventional charmonium state. This observation has stimulated a
number of distinct interpretations. These include a tetraquark
state~\cite{tetra}, $D\bar{D}^{\ast}$ molecule~\cite{molecule},
hadroquarkonium~\cite{hadron}, and other
configurations~\cite{other}. More recently, BESIII observed
another charged charmoniumlike state, $Z_c(4020)^{\pm}$, in $\EE
\to \pp h_c$~\cite{guo}. These states, together with similar
states observed in the bottomonium system~\cite{zbs}, indicate the
existence of  a new class of hadrons.

A strange partner of the $Z(3900)^{\pm}$, called $Z_{cs}$, may
exist in the above scenarios. The mass of a $J^P = 1^+$  $D_s
\bar{D}^{\ast}$ molecular state was first predicted~\cite{lee}
using QCD sum rules with $M(Z_{cs}) = (3.97 \pm 0.08)$~GeV/$c^2$,
which is very close to the $D_s^+ \bar{D}^{\ast0}$ threshold of
3.976~GeV/$c^2$. Using the same QCD sum rules, the authors of
Ref.~\cite{zcs} calculated the decay widths of the $Z_{cs}^+$ to
$K^+ \jpsi$, $K^{\ast+}\eta_c$, $D_s^+ \bar{D}^{\ast0}$ and
$\bar{D}^{0} D_s^{\ast+}$, assuming the $Z_{cs}$ to be a
tetraquark state. Such a state is also predicted in the
single-kaon emission model~\cite{liux}.

Using a data sample of 673~fb$^{-1}$ collected at or near
$\sqrt{s}=10.58$~GeV, Belle has observed abundant $\EE \to
\kkjpsi$ signal events via initial state radiation
(ISR)~\cite{belle-kkll}. In addition, there is one very broad
structure in the $\kkjpsi$ mass spectrum; fits using either a
single Breit-Wigner (BW) function, or the $\psi(4415)$ plus a
second BW function yield resonant parameters that are very
different from those of the currently tabulated excited $\psi$
states~\cite{PDG}. Unfortunately, the $M(K^\pm\jpsi)$ distribution
is not shown in Ref.~\cite{belle-kkll}.

In this paper, we report the updated measurement of the cross
sections for $\EE\to \kkjpsi$ and $\ksksjpsi$ between threshold
and 6.0~GeV/$c^2$ and examine possible resonant structures in the
cross section of the process $\EE\to \kkjpsi$ as well as in the
$K^{\pm}\jpsi$ and $\kk$ systems. The results are based on the
full Belle data sample with an integrated luminosity of
980~fb$^{-1}$ collected on or near the $\Upsilon(nS)$ ($n=$1, 2,
..., 5).

The Belle detector at the KEKB asymmetric-energy $\EE$
collider~\cite{kekb} is described in detail
elsewhere~\cite{belle-detector}. This analysis supersedes that
reported in Ref.~\cite{belle-kkll} where a subset of the Belle
data sample was used.

We use the {\sc phokhara}~\cite{phokahara} program to generate
signal Monte Carlo (MC) events and determine experimental
efficiencies. In the generator, one or two photons are allowed to
be emitted before forming the resonance $X$; then $X$ decays into
$\kkjpsi$ with $\jpsi$ decaying into $\ell^+\ell^-$ ($\ell=e$ or
$\mu$). When generating the MC sample, the mass of $X$ is fixed to
a certain value while the width is set to zero. In $X\to \kkjpsi$,
a pure $S$-wave between the $\kk$ system and the $\jpsi$ as well
as between the $K^+$ and $K^-$ is assumed. The invariant mass of
the $\kk$ system is generated uniformly in phase space. To
estimate the model uncertainty, we also generate events with a
$\kk$ invariant mass distributed like $M(\pp)$ in $\psi(2S) \to
\pp \jpsi$ decays~\cite{bai}.

\section{Event Selection}

The selection of $\kk\LL$ events is the same as in
Ref.~\cite{belle-kkll}. For the events of interest, we require
four well reconstructed charged tracks with zero net charge. The
impact parameters of these tracks perpendicular to and along the
beam direction with respect to the interaction point are required
to be less than 0.5~cm and 4.0~cm, respectively. For each charged
track, a combined likelihood ratio from various detector
subsystems is formed to identify different particle species
($e,\mu,\pi,K,p$). Tracks with $\mathcal{R}_K =
\frac{\mathcal{L}_K} {\mathcal{L}_K+\mathcal{L}_\pi}> 0.6$ are
identified as kaons with an efficiency of about 92\%; about 4\%
are misidentified pions~\cite{pid}. Similar ratios are also
defined for leptons~\cite{EID, MUID}. For electrons from $\jpsi\to
\EE$, one track should have $\mathcal{R}_e>0.95$ and the other
$\mathcal{R}_e>0.05$; for muons from $\jpsi\to \MM$, at least one
track should have $\mathcal{R}_\mu>0.95$; in cases where the other
has no muon identification ($\mathcal{R}_\mu=0$), in order to
suppress fake muon tracks, the cosine of the polar angle of each
muon track in the $\kk \MM$ CM frame is required to be less than
0.7. Events with $\gamma$ conversions are removed by requiring
$\mathcal{R}_e<0.75$ for the $\kk$ tracks. In $\jpsi\to \EE$,
$\gamma$ conversion events are further suppressed by requiring the
invariant mass of $\kk$ to be larger than 1.05~GeV/$c^2$; this
also removes the events with a $\phi$ meson in the final state. In
$\jpsi\to \MM$, the invariant mass of $\kk$ is required to be
outside  a $\pm 10$~MeV/$c^2$ interval around the $\phi$ nominal
mass to remove events with a $\phi$ meson in the final state,
possibly produced via $\EE \to \gamma \gamma^{\ast}
\gamma^{\ast}\to\gamma \phi \ell^+ \ell^-$. There is only one
combination of $\kk\LL$ in each event after the above event
selections.

\begin{figure*}[htbp]
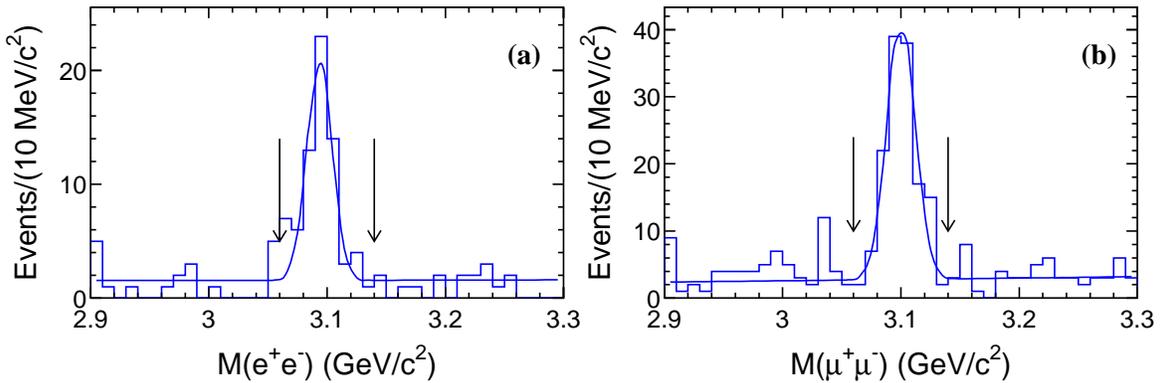

\psfig{file=fig1a.epsi,height=5cm} \put(-30,120){ \bf (a)}
\psfig{file=fig1b.epsi,height=5cm} \put(-30,120){ \bf (b)}
\caption{Invariant mass distributions of (a) $\EE$ and (b) $\MM$
for selected $\kk\LL$ candidates. The curves show the best fits to
the mass spectra, and the arrows show the required $\jpsi$ signal
regions defined in the text.} \label{mll}
\end{figure*}

The ISR events are identified by the requirement $|M^2_{\rm rec}|<
1.0$~(GeV/$c^2$)$^2$, where $M^2_{\rm rec} =
(P_{CM}-P_{K^+}-P_{K^-}-P_{\ell^+}-P_{\ell^-})^2$ and $P_i$
represents the four-momentum of the corresponding particle in the
$\EE$ CM frame. Clear $\jpsi$ signals are observed in both
$\jpsi\to \EE$ and $\jpsi\to \MM$ modes, as shown in
Fig.~\ref{mll}. We define the $\jpsi$ signal region as
$3.06<M(\LL)<3.14$~GeV/$c^2$ (with the mass resolution of lepton
pairs being about 17~MeV/$c^2$), and the $\jpsi$ mass sideband as
$2.91<M(\LL)<3.03$ or $3.17<M(\LL)<3.29$~GeV/$c^2$, which is three
times the width of the signal region. Here, final state radiation
and bremsstrahlung energy loss are recovered by adding the
four-momentum of photons detected within a $5^\circ$ cone around
the electron and positron direction in the $e^+ e^-$ invariant
mass calculation.

For selection of $\EE \to K_S^0 K_S^0 \LL$ events within the same
data sample, all the selection criteria are the same as for
$\kkjpsi$ except that the selection of $\kk$ is replaced by the
selection of two $K_S^0$. For a $K_S^0$ candidate decaying into
$\pp$, we require that the invariant mass of the $\pp$ pair lie
within a $\pm 11$~MeV/$c^2$ interval around the $K_S^0$ nominal
mass, which contains around 95\% of the signal according to the MC
simulation, and that the pion pair has a displaced vertex and
flight direction consistent with a $K_S^0$ originating from the
interaction point~\cite{ks}.

\begin{figure*}[htbp]
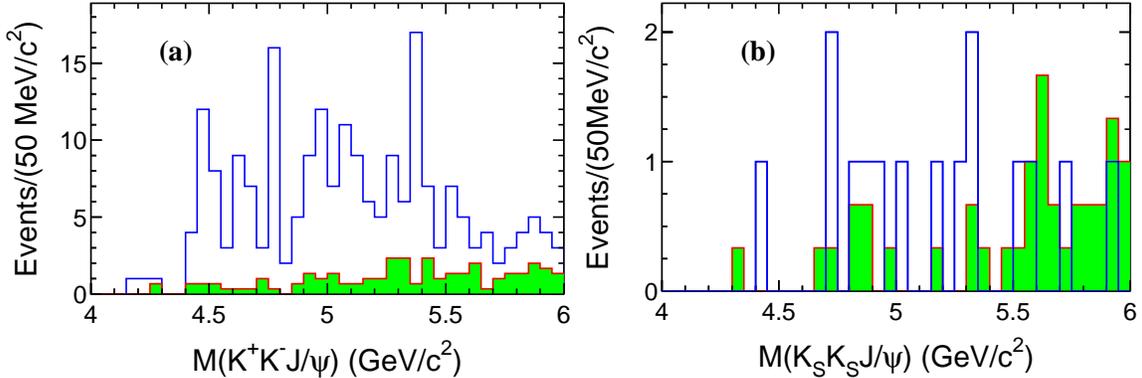

\includegraphics[height=5cm]{fig2a.epsi}
\includegraphics[height=5cm]{fig2b.epsi}
 \put(-370,120){\bf (a)}
 \put(-150,120){\bf (b)}
\caption{The invariant mass distributions of the $\kkjpsi$ (a) and
$\ksksjpsi$ (b) candidates. The open histograms are from the
$\jpsi$ signal region, while the shaded ones are from the $\jpsi$
mass sideband regions after a proper normalization.} \label{mass}
\end{figure*}

Figure~\ref{mass} shows the $\kk\LL$ and $K_S^0K_S^0\LL$ invariant
mass~\cite{def-mass} distributions after applying the above
selection, together with the backgrounds estimated from the
normalized $\jpsi$ mass sidebands. The $\kk\LL$  invariant mass
distribution is similar to that in Ref.~\cite{belle-kkll} and
shows a broad enhancement around 4.4-5.5~GeV/$c^2$. In addition,
there are 3 events near $\sqrt{s} = 4.26$~GeV. It is evident from
the figure that the background estimated from the $\jpsi$ mass
sidebands is low, which indicates that the background from the
non-$\jpsi$ final states is small. The other backgrounds not shown
in the sidebands include: (1) $\kkjpsi$ with $\jpsi$ decaying into
final states other than lepton pairs; (2) $X\jpsi$, with $X$ not
being $\kk$, such as $\pp$. The number of these background events
is found to be small from MC simulation and thus they are
neglected. Non-ISR production of the $\EE\to \kkjpsi$ process,
such as $\EE\to \gamma\gamma^*\gamma^*\to \gamma\phi\jpsi$, is
calculated to be small~\cite{zhuk} and is neglected. For the
$M(K_S^0 K_S^0 \jpsi)$ distribution, there are only 10 signal
candidate events between 4.6 and 5.5~GeV/$c^2$ with 4 background
events estimated from the $\jpsi$ mass sidebands. In other
regions, the number of events in the $\jpsi$ signal region is
about the same as expected from the normalized sideband events.

\begin{figure*}[htbp]
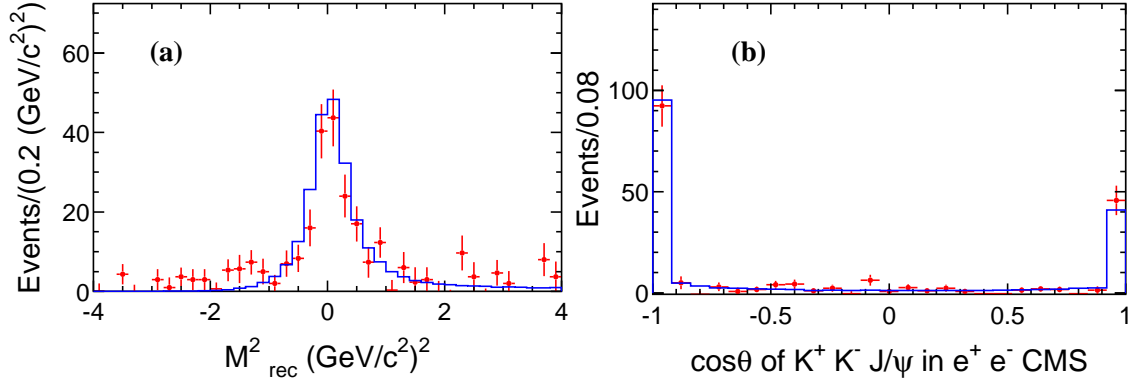

\includegraphics[height=5cm]{fig3a.epsi}
\includegraphics[height=5cm]{fig3b.epsi}
 \put(-370,120){\bf (a)}
 \put(-150,120){\bf (b)}
\caption{ (a) The distribution of the squared mass recoiling
against the $\kkjpsi$ system distribution and (b) the polar angle
distribution of the $\kkjpsi$ system in the $\EE$ CM frame for the
selected $\kkjpsi$ events with invariant masses between 4.0 and
6.0~GeV/$c^2$. The points with error bars are data with the
normalized $\jpsi$ mass sidebands subtracted; the solid histograms
are MC simulated events. } \label{isr}
\end{figure*}

Figures~\ref{isr}(a) and (b) show the distribution of the squared
mass recoiling against the $\kkjpsi$ system and the polar angle
distribution of the $\kkjpsi$ system in the $\EE$ CM frame,
respectively, for the selected $\kkjpsi$ events with invariant
masses between 4.0 and 6.0~GeV/$c^2$. The data, shown with the
normalized $\jpsi$ mass sidebands subtracted, agree well with the
MC simulation (open histograms), indicating the existence of
signals that are produced from ISR.

\section{Cross Sections}

\begin{figure}[htbp]
 \includegraphics[height=6cm]{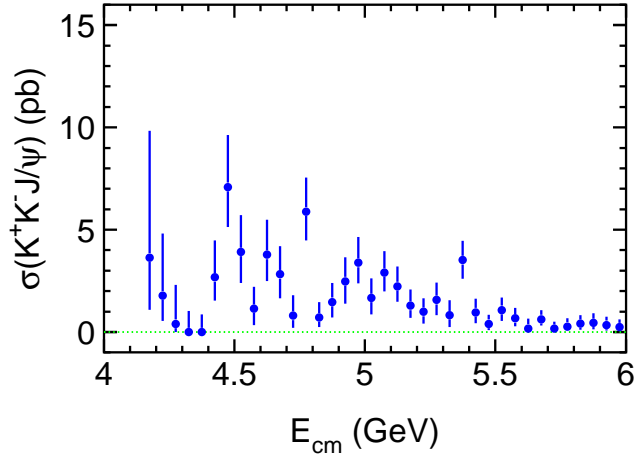}
\caption{The measured $\EE \to \kkjpsi$ cross sections for CM
energies up to 6.0~GeV (points with error bars). The errors are
statistical and are determined by the numbers of signal and
background events; a 7.8\% systematic error that is common for all
data points is not included.
} \label{cs}
\end{figure}

The $\EE\to\kkjpsi$ cross section at each energy point is
calculated using
 $$
 \sigma_i=\frac{n^{\rm obs}_i-f\times n^{\rm bkg}_i}
      {\mathcal{L}_i\cdot\epsilon_i\cdot\mathcal{B}(\jpsi\to\LL)},
 $$
where $n^{\rm obs}_i$, $n^{\rm bkg}_i$, $f$, $\epsilon_i$, and
$\mathcal{L}_i$ are the number of observed events in data, the
number of background events estimated from the $\jpsi$ sidebands,
the scale factor ($f=1/3$), the detection efficiency, and the
effective ISR luminosity obtained from the QED
calculation~\cite{kuraev} in the $i$-th energy bin, respectively;
$\mathcal{B}(\jpsi\to \LL)=11.87\%$ is taken from Ref.~\cite{PDG}.
According to the MC simulation, the efficiency for $K^+ K^-
J/\psi$ ($K_S^0 K_S^0 J/\psi$) increases smoothly from 1.69\%
(0.30\%) at 4.2~GeV/$c^2$, 7.53\% (0.56\%) at 4.6~GeV/$c^2$,
11.50\% (1.04\%) at 5.2~GeV/$c^2$, to 14.93\% (1.45\%) at
5.8~GeV/$c^2$. Figure~\ref{cs}
shows the measured cross sections for $\EE\to \kkjpsi$, where the
error bars indicate the combined statistical errors of the signal
and the background events, following the procedure in
Ref.~\cite{stat-err}.
The measured $\EE\to\kkjpsi$ cross sections are consistent with
previously published results~\cite{belle-kkll} with improved
precision. Similarly, the $\EE\to\ksksjpsi$ cross section is
calculated. Since the number of $\ksksjpsi$ signal events is very
small, we give an average cross
section for $\EE \to \ksksjpsi$ of $1.8\pm 0.6 (\rm stat.)$~pb
between 4.4 and 5.2~GeV/$c^2$. The result is consistent with the
previously published result of $1.8^{+1.4}_{-1.1} (\rm
stat.)$~pb~\cite{belle-kkll} with better precision.
Tables~\ref{xs_kkjpsi} and ~\ref{xs_ksksjpsi} list the final results and all the information
used in the cross section calculation for $\EE\to \kkjpsi$ and
$\ksksjpsi$, respectively.

\begin{table*}
\caption{Cross sections ($\sigma$) of $\EE\to \kkjpsi$. We also
list the $\EE$ center-of-mass energy ($\sqrt{s}$), number of
observed events ($n^{\rm obs}$), number of backgrounds estimated
from $\jpsi$ mass sidebands ($n^{\rm bkg}$), detection efficiency
($\epsilon$), and effective ISR luminosity ($\mathcal{L}$). All
values are calculated for 50~MeV bin size and $\sqrt{s}$ is the
central value of the bin. The first errors are statistical and the
second ones systematic. For the
bins with lower limit at zero at 68.3\% confidence interval, a
confidence interval is given with systematic error
included~\cite{stat-err}.}\label{xs_kkjpsi}
\begin{center}
\begin{tabular}{cccccc|cccccc}
 \hline
  $\sqrt{s}$ (GeV) & $n^{\rm obs}$ & $n^{\rm bkg}$ & $\epsilon$ (\%) &
  $\mathcal{L}$ (pb$^{-1}$) & $\sigma$ (pb) &
  $\sqrt{s}$ (GeV) & $n^{\rm obs}$ & $n^{\rm bkg}$ & $\epsilon$ (\%) &
  $\mathcal{L}$ (pb$^{-1}$) & $\sigma$ (pb)
  \\\hline
  4.175  & 1  & 0  & 1.12 & 207 & $3.6_{-2.6}^{+6.2}\pm0.3$ & 5.125  & 9  & 2 & 11.0 & 286 & $2.2_{-0.8}^{+1.0}\pm0.2$   \\
  4.225  & 1  & 0  & 2.25 & 210 &$1.8_{-1.3}^{+3.1}\pm0.2$  & 5.175  & 6  & 3 & 11.3 & 291 & $1.3_{-0.6}^{+0.8}\pm0.1$   \\
  4.275  & 1  & 2  & 3.24 & 214 & [0, 2.3]                  & 5.225  & 5  & 3 & 11.6 & 296 & $1.0_{-0.6}^{+0.7}\pm0.1$   \\
  4.325  & 0  & 0  & 4.13 & 218 & [0, 1.3]                  & 5.275  & 9  & 7 & 11.9 & 301 & $1.6_{-0.8}^{+0.9}\pm0.2$   \\
  4.375  & 0  & 0  & 4.91 & 221 & [0, 1.1]                  & 5.325  & 6  & 7 & 12.3 & 306 & $0.8_{-0.6}^{+0.8}\pm0.1$   \\
  4.425  & 4  & 0  & 5.61 & 225 &$2.7_{-1.2}^{+1.8}\pm 0.3$ & 5.375  & 17 & 2 & 12.6 & 311 & $3.5_{-1.0}^{+1.0}\pm0.3$   \\
  4.475  & 12 & 0  & 6.23 & 229 & $7.1_{-2.0}^{+2.6}\pm0.6$ & 5.425  & 7  & 7 & 12.9 & 317 & $1.0_{-0.6}^{+0.7}\pm0.1$   \\
  4.525  & 8  & 2  & 6.78 & 233 & $3.9_{-1.6}^{+1.8}\pm0.4$ & 5.475  & 3  & 3 & 13.2 & 323 & $0.4_{-0.3}^{+0.5}\pm0.1$   \\
  4.575  & 3  & 2  & 7.27 & 237 & $1.1_{-0.8}^{+1.1}\pm0.1$ & 5.525  & 7  & 4 & 13.5 & 328 & $1.1_{-0.6}^{+0.7}\pm0.1$   \\
  4.625  & 9  & 2  & 7.72 & 241 &$3.8_{-1.3}^{+1.8}\pm0.3$  & 5.575  & 5  & 4 & 13.8 & 334 & $0.7_{-0.4}^{+0.6}\pm0.1$   \\
  4.675  & 7  & 1  & 8.13 & 245 & $2.8_{-1.2}^{+1.4}\pm0.3$ & 5.625  & 3  & 6 & 14.1 & 340 & [0, 0.7]                    \\
  4.725  & 3  & 3  & 8.50 & 249 & $0.8_{-0.6}^{+1.0}\pm0.1$ & 5.675  & 4  & 1 & 14.4 & 346 & $0.6_{-0.3}^{+0.5}\pm0.1$   \\
  4.775  & 16 & 1  & 8.85 & 253 & $5.9_{-1.5}^{+1.7}\pm0.5$ & 5.725  & 2  & 3 & 14.6 & 353 & $0.2_{-0.2}^{+0.4}\pm0.1$  \\
  4.825  & 2 & 0  & 9.18  & 258 & $0.7_{-0.5}^{+0.8}\pm0.1$ & 5.775  & 3  & 4 & 14.8 & 359 & $0.3_{-0.3}^{+0.5}\pm0.1$   \\
  4.875  & 5 & 2  & 9.49  & 262 & $1.5_{-0.8}^{+1.0}\pm0.2$ & 5.825  & 4  & 4 & 15.0 & 366 & $0.4_{-0.3}^{+0.4}\pm0.1$   \\
  4.925  & 9 & 4  & 9.80  & 267 & $2.5_{-1.1}^{+1.2}\pm0.2$ & 5.875  & 5  & 6 & 15.1 & 373 & $0.5_{-0.4}^{+0.5}\pm0.1$  \\
  4.975  & 12& 3  & 10.1  & 271 & $3.4_{-1.1}^{+1.3}\pm0.3$ & 5.925  & 4  & 5 & 15.2 & 380 & $0.3_{-0.3}^{+0.5}\pm0.1$   \\
  5.025  & 7 & 4  & 10.4  & 276 & $1.7_{-0.9}^{+1.0}\pm0.2$ & 5.975  & 3  & 4 & 15.3 & 387 & $0.2_{-0.2}^{+0.4}\pm0.1$   \\
  5.075  & 11& 2  & 10.7  & 281 & $2.9_{-1.0}^{+1.1}\pm0.3$ &        &    &   &      &     &                             \\
  \hline
\end{tabular}
\end{center}
\end{table*}

\begin{table*}
\caption{Cross sections ($\sigma$) of $\EE\to \ksksjpsi$. We also
list the $\EE$ center-of-mass energy ($\sqrt{s}$), number of
observed events ($n^{\rm obs}$), number of backgrounds estimated
from $\jpsi$ mass sidebands ($n^{\rm bkg}$), detection efficiency
($\epsilon$), and effective ISR luminosity ($\mathcal{L}$). All
values are calculated for 50~MeV bin size and $\sqrt{s}$ is the
central value of the bin. As the
number of $\ksksjpsi$ signal events is small, a 68.3\% confidence
interval for the measured cross section is given with systematic
error included~\cite{stat-err}.}\label{xs_ksksjpsi}
\begin{center}
\begin{tabular}{cccccc|cccccc}
 \hline
  $\sqrt{s}$ (GeV) & $n^{\rm obs}$ & $n^{\rm bkg}$ & $\epsilon$ (\%) &
  $\mathcal{L}$ (pb$^{-1}$) & $\sigma$ (pb) &
  $\sqrt{s}$ (GeV) & $n^{\rm obs}$ & $n^{\rm bkg}$ & $\epsilon$ (\%) &
  $\mathcal{L}$ (pb$^{-1}$) & $\sigma$ (pb)
  \\\hline
  4.175  & 0  & 0  & 0.26 & 207 & [0, 21] &  5.125  & 0  & 0  & 0.96 & 286 & [0, 4.0]  \\
  4.225  & 0  & 0  & 0.30 & 210 & [0, 18] &  5.175  & 1  & 1  & 1.00 & 291 & [0, 6.7]    \\
  4.275  & 0  & 0  & 0.34 & 214 & [0, 16] &  5.225  & 0  & 0  & 1.03 & 296 & [0, 3.6] \\
  4.325  & 0  & 1  & 0.37 & 218 & [0, 10] &  5.275  & 1  & 0  & 1.07 & 301 & [0.7, 7.1]  \\
  4.375  & 0  & 0  & 0.41 & 221 & [0, 13] &  5.325  & 2  & 2  & 1.11 & 306 & [0.7, 8.7] \\
  4.425  & 1  & 0  & 0.44 & 225 & [2.5, 23]&  5.375  & 0  & 1  & 1.15 & 311 & [0, 2.2]  \\
  4.475  & 0  & 0  & 0.48 & 229 & [0, 10] &  5.425  & 0  & 0  & 1.18 & 317 & [0, 3.0] \\
  4.525  & 0  & 0  & 0.52 & 233 & [0, 9.2]&  5.475  & 0  & 1  & 1.22 & 323 & [0, 2.0]  \\
  4.575  & 0  & 0  & 0.55 & 237 & [0, 8.4]&  5.525  & 1  & 1  & 1.26 & 328 & [0, 4.7]  \\
  4.625  & 0  & 0  & 0.59 & 241 & [0, 7.8] & 5.575  & 1  & 3  & 1.30 & 334 & [0, 3.4]  \\
  4.675  & 0  & 1  & 0.63 & 245 & [0, 5.0] & 5.625  & 0  & 5  & 1.34 & 340 & [0, 1.0]  \\
  4.725  & 2  & 1  & 0.66 & 249 & [2.5, 20]& 5.675  & 0  & 2  & 1.37 & 346 & [0, 1.3]   \\
  4.775  & 0  & 0  & 0.70 & 253 & [0, 6.2] &  5.725  & 1  & 1  & 1.41 & 353 & [0, 3.9]  \\
  4.825  & 1  & 2  & 0.74 & 258 & [0, 8.5] &  5.775  & 0  & 2 & 1.45 & 359 &  [0, 1.2] \\
  4.875  & 1  & 2  & 0.77 & 262 & [0, 7.9] &  5.825  & 0  & 2 & 1.49 & 366 &  [0, 1.1] \\
  4.925  & 1  & 0  & 0.81 & 267 & [1.1, 11]     &  5.875  & 0  & 2 & 1.53 & 373 &  [0, 1.1] \\
  4.975  & 0  & 1  & 0.85 & 271 & [0, 3.3] &  5.925  & 1  & 4 & 1.56 & 380 &  [0, 2.5] \\
  5.025  & 1  & 0  & 0.89 & 276 & [1.0, 9.4]      &  5.975  & 0  & 3 & 1.60 & 387 &  [0, 0.7] \\
  5.075  & 0  & 0  & 0.92 & 281 & [0, 4.3] &    &  &  &  &  &   \\
  \hline
\end{tabular}
\end{center}
\end{table*}

Systematic error sources and their contributions in the cross
section measurements are summarized in Table~\ref{sys-err}. The
lepton pair identification uncertainties, measured from a pure
control sample of $\EE \to \gamma_{\rm ISR} \psip$ events with
$\psip\to \ppjpsi$, $\jpsi\to \ell^+ \ell^-$, are 3.5\% and 1.8\%
for $\EE$ and $\MM$, respectively~\cite{Belle-z}. The uncertainty
due to kaon particle identification is 1.2\% for each kaon.
Tracking efficiency uncertainties are estimated to be 1.3\% per
kaon track and 0.35\% per lepton track, which are fully correlated
in the momentum and angle regions of interest for signal events.
The systematic uncertainty in the $K_S^0$ reconstruction
efficiency is estimated by using the control samples of
reconstructed $D^{\ast \pm}$ decays with the decay chain $D^{\ast
\pm} \to \pi_s^{\pm} D^0$, $D^0 \to K_S^0 \pp$. We find that the
MC efficiency is higher than in data by $(2.1\pm 0.7)\%$. We take
2.8\% as the systematic uncertainty for each $K_S^0$ selection.
The uncertainties associated with the $\jpsi$ mass window and
$|M^2_{\rm rec}|$ requirements are also estimated using pure
$\psip\to \ppjpsi$ events. It is found that MC efficiencies are
higher than in data by ($4.5\pm 0.4$)\% in the $\EE$ mode and
$(4.1\pm 0.2)$\% in the $\MM$ mode. The differences in
efficiencies are corrected and the uncertainties in the correction
factors are taken as systematic errors. They contribute 0.6\% for
the $\EE$ and 0.3\% for the $\MM$ mode in total for the $\jpsi$
mass window together with the $|M^2_{\rm rec}|$
requirements~\cite{Belle-z}. Estimating the backgrounds using
different $\jpsi$ mass sidebands results in a change of background
events at the 0.12/50~MeV/$c^2$ level for $\kkjpsi$ and
0.008/50~MeV/$c^2$ level for $\ksksjpsi$, corresponding to an
average change of about 2.6\%  for $\kkjpsi$ and 14\% for
$\ksksjpsi$ in the cross section. Belle measures the total
luminosity with a precision of 1.4\% using Bhabha events. The {\sc
phokhara} generator calculates the ISR photon radiator function
with 0.1\% accuracy~\cite{phokahara}. The dominant uncertainties
due to the generator come from the three-body decay dynamics;
there is no good model to describe the $\kk$ mass spectrum.
Simulations with modified $\kk$ invariant mass distributions such
as $M(\pp)$ in $\psi(2S) \to \pp \jpsi$~\cite{bai} yield
efficiencies that are higher by 3.3-4.8\% for different $\kkjpsi$
masses. We take 4.8\% as a conservative estimation for the
$\kkjpsi$ mass values. Similarly, we take 4.5\% for the
$\ksksjpsi$ mode.  The angular distributions of the final state
particles for  selected $\kkjpsi$ events from data are consistent
with the MC simulations and no evidence is found for non-S-wave
components. The selected data sample contains at least four
charged tracks and so the trigger efficiency is higher than 98\%
according to MC simulation. A 1.0\% systematic error is assigned
for the trigger uncertainty. The uncertainty of $\BR(\jpsi\to
\LL)=\BR(\jpsi\to \EE)+ \BR(\jpsi\to \MM)$ is taken as 1.0\% from
Ref.~\cite{PDG}. The uncertainty of $\BR(K_S^0 \to \pp)$ is
neglected. Finally, the MC statistical error on the efficiency is
1.5\%. Assuming that all the sources are independent and adding
them in quadrature, we obtain a total systematic error on the
cross section of 7.8\% for the $\kkjpsi$ and 16\% for $\ksksjpsi$
final states.

\begin{table}[htbp]
\caption{Systematic errors in $\EE\to \kkjpsi$ and $\ksksjpsi$
cross section measurements.} \label{sys-err}
\begin{tabular}{c c c}
\hline\hline
 Source & $\kkjpsi$~(\%) & $\ksksjpsi$~(\%) \\\hline
 Particle identification &  3.6 & 2.6 \\
 Tracking & 3.3 & 0.7 \\
 $K_S^0$ selection & --- & 5.6 \\
 $\jpsi$ mass and $M^2_{\rm rec}$ selection & 0.4 & 0.4 \\
 Background estimation & 2.6 & 14 \\
 Integrated luminosity & 1.4 & 1.4 \\
 Generator & 4.8 & 4.5 \\
 Trigger efficiency & 1.0 & 1.0 \\
 Branching fractions & 1.0 & 1.0 \\
 MC statistics & 1.5 & 1.5 \\
 \hline
 Sum in quadrature & 7.8 & 16 \\
 \hline\hline
\end{tabular}
\end{table}


\section{Resonant Structures}

\begin{figure*}[htbp]
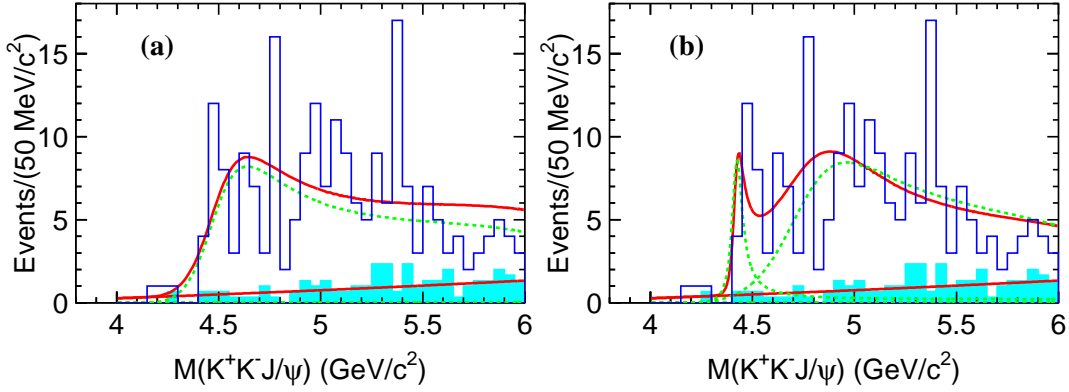

\includegraphics[height=7cm,angle=-90]{fig5a.epsi}
\includegraphics[height=7cm,angle=-90]{fig5b.epsi}
 \put(-350,-20){\bf (a)}
 \put(-150,-20){\bf (b)}
\caption{Fits to the $\kkjpsi$ invariant mass distribution. The
open histograms are the selected data in the $\jpsi$ signal region
while the shaded histograms show the normalized $\jpsi$ sideband
events. The solid curves show the best fit to the data and
sideband background with one BW function (a) and the coherent sum
of a BW function and the $\psi(4415)$ component (b).} \label{fit}
\end{figure*}

The unbinned maximum likelihood fit performed in
Ref.~\cite{belle-kkll} is  applied to the $\kkjpsi$ mass spectrum
in Fig.~\ref{mass}(a). The theoretical shape is multiplied by the
efficiency and effective luminosity, which are functions of the
$\kkjpsi$ invariant mass. The BW function for a spin-1 resonance
decaying into a final state $f$ with mass $M$, total width
$\Gamma_{\rm tot}$ and partial width $\Gamma_{\EE}$ to $\EE$ is
 $$
\sigma(s)=\frac{M^2}{s}\frac{12\pi\Gamma_{\EE}\BR(R\to
f)\Gamma_{\rm tot}} {(s-M^2)^2+M^2\Gamma_{\rm
tot}^2}\frac{\rho(\sqrt{s})}{\rho(M)},
 $$
where $\BR(R\to f)$ is the branching fraction of the resonance to
the final state $f$, and $\rho(\sqrt{s})$ is the three-body decay
phase space factor for $X\to \kk\jpsi$. We attempt to fit the
$\kkjpsi$ invariant mass spectrum using two different
parameterizations of the signal shape: (1) a single BW function
plus a background term; (2) a coherent sum of a BW function and a
$\psi(4415)$ component with mass and width fixed at their world
average values~\cite{PDG} plus a background term. The fit results
are shown in Fig.~\ref{fit}. The results of the fit for the BW
parameters [$M=(4482\pm 45)$~MeV/$c^2$, $\Gamma_{\rm tot}=(432\pm
56)$~MeV for model~(a) and $M=(4747\pm 117)$~MeV/$c^2$,
$\Gamma_{\rm tot}=(671\pm 86)$~MeV/$c^2$ for model~(b)] are
consistent with the previously published results within about
$2\sigma$, but the goodness of the fit ($\chi^2/ndf = 39/13 =3.0$
for model~(a) and $\chi^2/ndf=30/11=2.7$ for model~(b)) is
marginal. (Here, {\it ndf} is the number of degrees of freedom).
Thus, our two models can not describe the data well with the
increased statistics. Here, in determining the goodness of each
fit, we bin the data so that the expected number of events in a
bin is at least seven. Adding a coherent $Y(4260)$ amplitude in
the fit with mass and width fixed at the latest Belle
measurement~\cite{Belle-z} yields an upper limit on
$\BR(Y(4260)\to \kk\jpsi) \Gamma(Y(4260)\to \EE) < 1.7$~eV/$c^2$
at 90\% confidence level. A similar fit to the $\ksksjpsi$
invariant mass spectrum with a $Y(4260)$ amplitude yields an upper
limit on $\BR(Y(4260)\to K_S^0 K_S^0 \jpsi) \Gamma(Y(4260)\to \EE)
< 0.85$~eV/$c^2$ at 90\% confidence level.

\begin{figure*}[htbp]
\includegraphics[height=5.5cm]{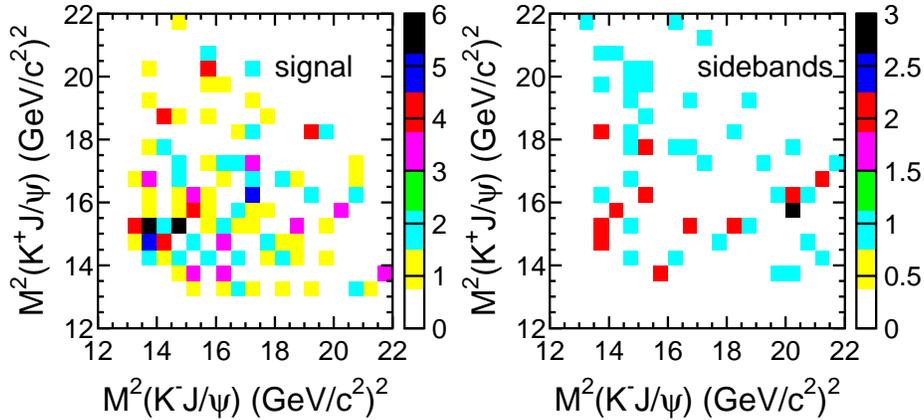}
\caption{Dalitz plots for the selected $\kkjpsi$ events for
$4.4<M(\kkjpsi)< 5.5$~GeV/$c^2$. The left panel is for events in
the $\jpsi$ signal region while the right is from the $\jpsi$ mass
sidebands (not normalized). } \label{dalitz}
\end{figure*}

\begin{figure*}[htbp]
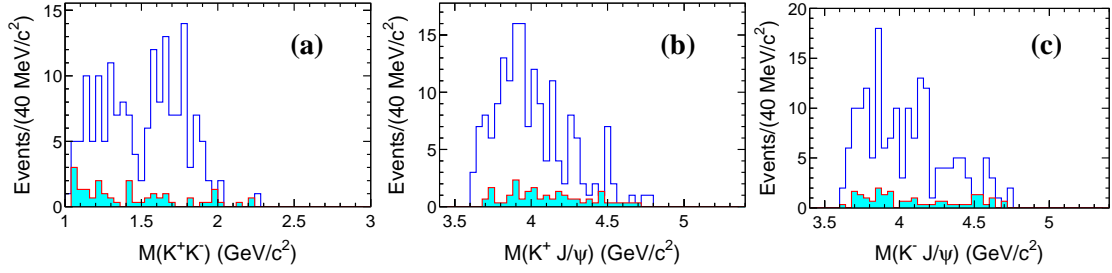

\includegraphics[height=3.5cm]{fig7a.epsi}
\includegraphics[height=3.5cm]{fig7b.epsi}
\includegraphics[height=3.5cm]{fig7c.epsi}
 \put(-310,80){\bf (a)}
 \put(-170,80){\bf (b)}
 \put(-30,80){\bf (c)}
\caption{Invariant mass distributions of (a) $\kk$, (b) $K^+
\jpsi$, and (c) $K^- \jpsi$ for $\kk\jpsi$ events with
$4.4<M(\kkjpsi)< 5.5$~GeV/$c^2$. Solid histograms are for events
in the $\jpsi$ signal region, and the shaded histograms are
normalized background from the $\jpsi$ mass sidebands.}
\label{projections}
\end{figure*}

Possible intermediate states are studied by examining the Dalitz
plot of the selected $\kkjpsi$ candidate events.
Figure~\ref{dalitz} shows the Dalitz plots of events in the
$\jpsi$ signal region and $\jpsi$ mass sidebands.
Figure~\ref{projections} shows a projection of the $M(\kk)$,
$M(K^+\jpsi)$, and $M(K^-\jpsi)$ invariant mass distributions.
Background events estimated from the normalized $\jpsi$ mass
sidebands are shown as the shaded histograms. No obvious
structures are observed in the $K^{\pm}\jpsi$ system.
The low statistics prevent us from extracting additional
information on the three-body dynamics.

\section{Summary}

The cross sections of $\EE\to \kkjpsi$ and $\ksksjpsi$
are measured from threshold to 6.0~GeV using the full Belle data
sample. There are clear $\kkjpsi$ signal events; however fits that
were tried before~\cite{belle-kkll}  with a smaller data set using
either a single BW function or using the $\psi(4415)$ plus a
second BW function are inadequate for the full data sample.
Possible intermediate states for the selected $\kkjpsi$ events are
also investigated by examining the Dalitz plot but no clear
structure is observed in the $K^{\pm} \jpsi$ system. Since there
are only a few $\ksksjpsi$ signal events and no structure is
observed in the $\ksksjpsi$ mass spectrum, the Dalitz plot of
$\ksksjpsi$ events is not examined. A larger data sample is
necessary to obtain more information about possible structures in
the $\kkjpsi$, $\ksksjpsi$, $\kk$ and $K^{\pm} \jpsi$ systems.


\acknowledgments

We thank the KEKB group for excellent operation of the
accelerator; the KEK cryogenics group for efficient solenoid
operations; and the KEK computer group, the NII, and PNNL/EMSL for
valuable computing and SINET4 network support. We acknowledge
support from MEXT, JSPS and Nagoya's TLPRC (Japan); ARC and DIISR
(Australia); FWF (Austria); NSFC (China); MSMT (Czechia); CZF,
DFG, and VS (Germany); DST (India); INFN (Italy); MOE, MSIP, NRF,
GSDC of KISTI, BK21Plus, and WCU (Korea); MNiSW and NCN (Poland);
MES and RFAAE (Russia); ARRS (Slovenia); IKERBASQUE and UPV/EHU
(Spain); SNSF (Switzerland); NSC and MOE (Taiwan); and DOE and NSF
(USA).




\begin{thebibliography}{**}
\bibitem{def-mass} In this paper, $M(\kk\LL)-M(\LL)+m_{\jpsi}$ is used instead of the invariant mass of the four
final state particles to improve the mass resolution. Here, $m_{\jpsi}$  is the nominal mass of the $\jpsi$ meson.
The same for the $M(K_S^0K_S^0\LL)$ calculation.
\end{thebibliography}
\end{document}